\documentclass[12pt]{article}

\usepackage{sbc-template}
\usepackage{graphicx,url}
\usepackage[utf8]{inputenc}  
\usepackage{subfiles}

\usepackage[abbreviations]{glossaries-extra}
\makeglossaries
\newcommand{\firstuseformat}[1]{#1}
\renewcommand{\glslinkpresetkeys}{
  \ifglsused{\glslabel}%
  {\letcs\glstextformat{@firstofone}}%
  {\let\glstextformat\firstuseformat}%
}
\newabbreviation{SE}{SE}{Software Engineering}
\newabbreviation{SEE}{SEE}{Software Engineering Education}
\newabbreviation{DAPP}{dApp}{Decentralized Application}
\newabbreviation{METASEE}{MetaSEE}{Metaverse-based Software Engineering Education}
\newabbreviation{XRAPP}{XR app}{XR application}
\newabbreviation{iL}{iL}{Immersive Learning}
\newabbreviation{MULSEMEDIA}{MulSeMedia}{Multiple Sensory Media}
\newabbreviation{AI}{AI}{Artificial Intelligence}
\newabbreviation{MR}{MR}{Mixed Reality}
\newabbreviation{XR}{XR}{eXtended Reality}
\newabbreviation{IDE}{IDE}{Integrated Development Environment}
\newabbreviation{UML}{UML}{Unified Modeling Language}
\newabbreviation{LA}{LA}{Learning Analytics}
\newabbreviation{SES}{SES}{Software Engineering Skill}
\newabbreviation{POC}{PoC}{Proof of Concept}
\newabbreviation{JS}{JS}{JavaScript}
\newabbreviation{VW}{VW}{Virtual World}
\newabbreviation{P2P}{P2P}{Peer-to-Peer}

\usepackage{listings, xcolor}

\definecolor{verylightgray}{rgb}{.97,.97,.97}

\lstdefinelanguage{Solidity}{
	keywords=[1]{anonymous, assembly, assert, balance, break, call, callcode, case, catch, class, constant, continue, constructor, contract, debugger, default, delegatecall, delete, do, else, emit, event, experimental, export, external, false, finally, for, function, gas, if, implements, import, in, indexed, instanceof, interface, internal, is, length, library, log0, log1, log2, log3, log4, memory, modifier, new, payable, pragma, private, protected, public, pure, push, require, return, returns, revert, selfdestruct, send, solidity, storage, struct, suicide, super, switch, then, this, throw, transfer, true, try, typeof, using, value, view, while, with, addmod, ecrecover, keccak256, mulmod, ripemd160, sha256, sha3}, 
	keywordstyle=[1]\color{blue}\bfseries,
	keywords=[2]{address, bool, byte, bytes, bytes1, bytes2, bytes3, bytes4, bytes5, bytes6, bytes7, bytes8, bytes9, bytes10, bytes11, bytes12, bytes13, bytes14, bytes15, bytes16, bytes17, bytes18, bytes19, bytes20, bytes21, bytes22, bytes23, bytes24, bytes25, bytes26, bytes27, bytes28, bytes29, bytes30, bytes31, bytes32, enum, int, int8, int16, int24, int32, int40, int48, int56, int64, int72, int80, int88, int96, int104, int112, int120, int128, int136, int144, int152, int160, int168, int176, int184, int192, int200, int208, int216, int224, int232, int240, int248, int256, mapping, string, uint, uint8, uint16, uint24, uint32, uint40, uint48, uint56, uint64, uint72, uint80, uint88, uint96, uint104, uint112, uint120, uint128, uint136, uint144, uint152, uint160, uint168, uint176, uint184, uint192, uint200, uint208, uint216, uint224, uint232, uint240, uint248, uint256, var, void, ether, finney, szabo, wei, days, hours, minutes, seconds, weeks, years},	
	keywordstyle=[2]\color{teal}\bfseries,
	keywords=[3]{block, blockhash, coinbase, difficulty, gaslimit, number, timestamp, msg, data, gas, sender, sig, value, now, tx, gasprice, origin},	
	keywordstyle=[3]\color{violet}\bfseries,
	identifierstyle=\color{black},
	sensitive=true,
	comment=[l]{//},
	morecomment=[s]{/*}{*/},
	commentstyle=\color{gray}\ttfamily,
	stringstyle=\color{red}\ttfamily,
	morestring=[b]',
	morestring=[b]"
}

\title{Towards a Blockchain-based Software Engineering Education}

\author{Filipe Fernandes\inst{1}\inst{2}, Cláudia Werner\inst{1}}

\address{Federal University of Rio de Janeiro -- COPPE/UFRJ\\
    Systems Engineering and Computer Science Department\\
    Caixa Postal 68.511 -- 21.941-972 -- Rio de Janeiro -- RJ -- Brazil
    \nextinstitute
    Federal Institute of Southeast Minas Gerais -- IF Sudeste MG\\
    Computer Science Department\\
    Caixa Postal 413 -- 36.909-300 -- Manhuaçu -- MG -- Brazil
    \email{\{ffernandes,werner\}@cos.ufrj.br}
}

\begin{document} 

\maketitle

\begin{abstract}
Blockchain technologies for rewards in education are gaining attraction as a promising approach to motivate student learning and promote academic achievement. By providing tangible rewards for educational attainment and engagement, such as digital tokens, educators can motivate learners to take a more active role in their learning and increase their sense of ownership and responsibility for their academic outcomes. In this context, this work proposes the Software Engineering Skill (SES) token as a way of rewarding students in order to improve their experiences in Software Engineering Education (SEE). We performed a proof of concept and conclude that SES token can be deployed in a platform to support SEE.
\end{abstract}
     

\section{Introduction} \label{sec:introduction}

Blockchain technology in the field of education is currently in its nascent stage, with only a limited number of educational institutions adopting this technology. This technology could disrupt the traditional role of educational institutions as certification agents, and offer students greater access to learning opportunities \cite{Nespor2019}.
According to \cite{Tan2022}, blockchain technology in education can provide equal opportunities for students to develop themselves through the implementation of a token economy in different classroom settings. 
In educational contexts, the adoption of a token economy is perceived as an essential mechanism for sustaining interest and engagement in the educational process.

For this reason, this work proposes the \gls{SES} token as an approach to support blockchain-based \gls{SEE}. \gls{SES} token aims to engage and motivate students and developers in order to improve experiences in \gls{SEE}, as well as increase developer contributions on the educational platform providing \gls{SES} tokens.

This paper is organized as follows: Section \ref{sec:context} presents the research context of which this work is part. Our proposal is described in Section \ref{sec:approach} and a proof of concept is presented in Section \ref{sec:proof-of-concept}. Finally, our conclusions and future directions are described in Section \ref{sex:conclusion}.

\section{Research Context} \label{sec:context}

This work is an extension of a doctoral thesis in progress that proposes an approach to enable the Metaverse to support both \glspl{XRAPP} development based on software reuse techniques and mechanisms to improve learning outcomes in \gls{SE}, allowing educators and students to have immersive experiences and increasing adoption of \gls{iL} in \gls{SEE} \cite{FERNANDES@2023MetaSEE}. \gls{METASEE} approach aims to provide an interoperable and scalable structure composed of the Metaverse's main concepts and technologies grouped in five layers: Physical, Virtual, Metaverse Engine, MetaSEE, and Infrastructure.



\textit{Physical Layer} corresponds to the main entities external to the Metaverse that belong to the physical and real world.
\textit{Virtual Layer} establishes the main components of the ``virtualization'' of physical layer elements. 
\textit{Metaverse Engine Layer} is composed of general \textit{Technologies}, as well as \textit{Economics} and \textit{Security} of the Metaverse.
\textit{MetaSEE Layer} is the main contribution to support \gls{SEE} through the Metaverse. \textit{Development Tools} component should provide a set of mechanisms to facilitate the development of \glspl{XRAPP} for \gls{SEE} considering the range of complexity and characteristics involved.
\textit{Integration Tools} component should provide reusable functionality for \gls{SEE}.
\textit{\gls{LA}} is the component that must guarantee the maintenance of the learning performance of the Metaverse for \gls{SEE} users.
\textit{Infrastructure Layer} deals with network and decentralization aspects of the Metaverse.

As one of the results, a platform has been developed in order to validate and evaluate the proposal. In general, \gls{METASEE} platform is composed of a web application and \glspl{XRAPP}. From the web application, students can search and access \glspl{VW} created by educators and other students. These \glspl{VW} are mainly composed of \gls{SE}-specific features, such as \gls{UML} diagrams, code editing, importing repositories, management tools, etc. There is an architecture to support developers' contributions to add \gls{SE}-specific features (extensions) to the platform. 

In order to contribute to the \gls{METASEE} approach, more specifically in the \textit{Decentralized Storage} component of \textit{Infrastructure Layer}, this work proposes to use blockchain technology to create a tokenization mechanism, which will be detailed in Section \ref{sec:approach}.

\section{Blockchain-based Software Engineering Education} \label{sec:approach}

The main goal of this work is to propose a solution that meets the \textit{Decentralized Storage} component in order to contribute to the \gls{METASEE} approach.
This work designs a mechanism through tokenization with the purpose of fostering a digital reward in order to improve student engagement, as well as motivating developers to contribute extensions (plugins) to the \gls{METASEE} approach.

Therefore, we defined the \gls{SES} as a \gls{METASEE} token to engage and motivate students and developers.
Considering the \gls{METASEE} approach context, \gls{SES} token can provide a mechanism based on ``learn to earn'' for students \cite{Tan2022}. 

The use of \gls{SES} token can provide tangible incentives for students, which can be used as rewards for achieving learning goals or performing other activities within the platform. In addition, creating a reward system can help making the learning process more playful and engaging, encouraging students to dedicate themselves more to their studies.

\gls{SES} token can also help make the platform more interactive and collaborative. For example, students can be rewarded for helping other students solve problems or answer questions. This can create a more engaged and collaborative community of students, which can be an additional source of motivation for learning. Additionally, \gls{SES} token can be used as a way to record and recognize student achievements. For example, successful completion of an activity can result in a token being issued, which can be used to unlock additional content or access special features. This can help create a sense of accomplishment and progress for students, which can motivate them to keep learning and progressing.

Another way and use of the \gls{SES} token is in the gamification of the learning experience, making the \gls{METASEE} platform more attractive and engaging. By setting clear goals and rewards for students, \gls{SES} token can help create a more exciting and challenging learning environment that can keep students motivated and engaged longer.

For developers, \gls{SES} token can be an effective mechanism to motivate developers to contribute to the \gls{METASEE} platform. Developers can receive tokens as a reward for contributing new extensions and bug fixes. These rewards can provide a form of recognition for work done and can be used to encourage a culture of collaboration in the developer community. This can result in increased participation as well as increased \gls{SE}-specific features through extensions.
\section{Proof of Concept} \label{sec:proof-of-concept}

In this section is described the \gls{SES} token \gls{POC}.
Our implementation was developed with \gls{JS} and Solidity\footnote{https://soliditylang.org/} languages, as well as Hardhat\footnote{https://hardhat.org/} framework. Solidity is a smart contract programming language used on the Ethereum platform and Hardhat is a software development environment for Ethereum that allows testing, compiling, and deploying smart contracts. Hardhat is a useful tool for developers working with Solidity as it simplifies and automates many common smart contract development tasks.

\begin{figure}[ht]
    \centering
    \includegraphics[scale=.55]{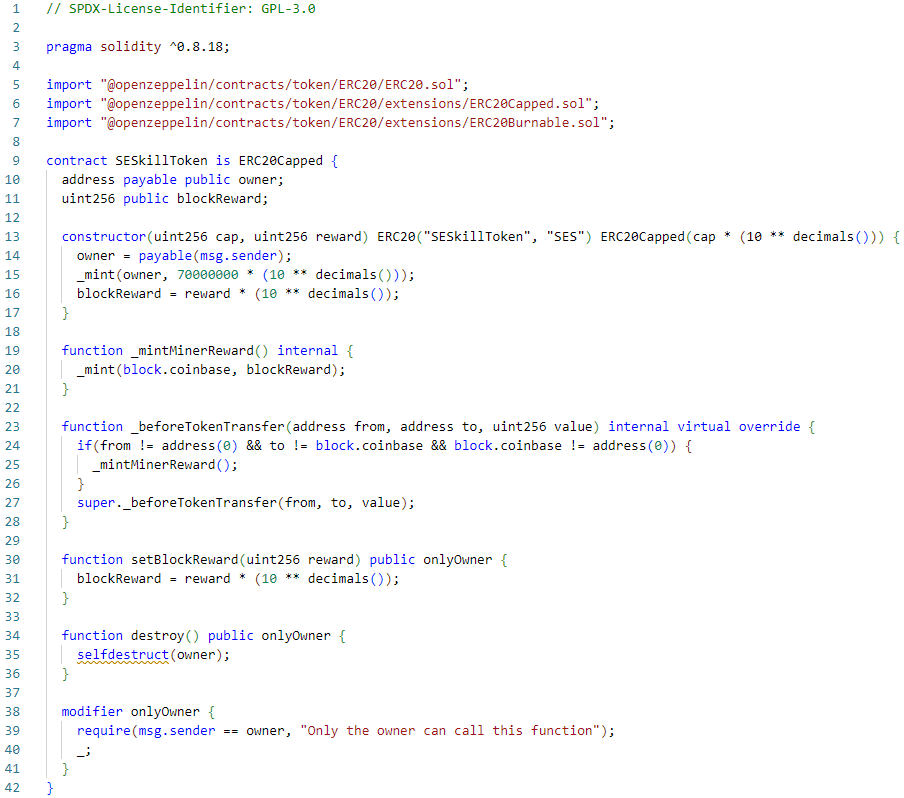}
    \caption{SES token code}
    \label{fig:ses-code}
\end{figure}

In order to prove a \gls{SES} token basic transaction between accounts, we implemented a smart contract written in Solidity that implements a custom ERC20 token called \textit{SESkillToken}, as shown in Figure \ref{fig:ses-code}. ERC20 defines a set of interfaces and standards for implementing digital tokens on the blockchain. This means that \textit{SESkillToken}, being a contract that implements the ERC20 standard, follows ERC20 interfaces and standards. This ensures that \textit{SESkillToken} is compatible with other tokens and applications that also implement the ERC20 standard. Lines 5 to 7 import contract inherits functionality from other OpenZeppelin\footnote{https://www.openzeppelin.com/} (open-source repository with standard contracts) contracts such as \textit{ERC20}, \textit{ERC20Capped} and \textit{ERC20Burnable}, which means that the \textit{SESkillToken} has the same functionality as those contracts, in addition to the custom functionality defined in this contract.

Line 13 defines a constructor that establishes the token's name and symbol, as well as its bid limit. In addition, it defines the contract owner's address and the miner reward amount per block. The \textit{\textunderscore mintMinerReward()} function (line 19) is responsible for issuing new coins to the miner who mined the current block, with a value equals to the one defined in the \textit{blockReward} variable. On line 23, \textit{\textunderscore beforeTokenTransfer()} function is a hook that runs before each token transfer. In this case, it checks if the transfer is taking place between an address different from the address of the block where the transfer is taking place. If this is true, then it calls the \textit{\textunderscore mintMinerReward()} function to issue the miner reward.

The contract also has other functions such as \textit{setBlockReward()} which allows the contract owner to set the miner reward amount per block and the \textit{destroy()} function which allows the owner to destroy the contract and recover the remaining funds. Finally, the contract uses an \textit{onlyOwner} modifier to ensure that only the owner of the contract can call functions that require owner privileges, such as \textit{setBlockReward()} and \textit{destroy()}.

\begin{figure}
    \centering
    \includegraphics[scale=.55]{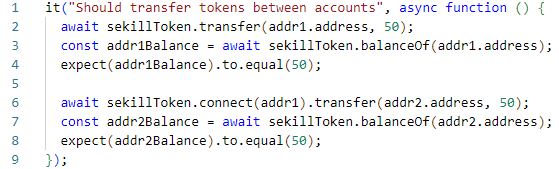}
    \caption{SES token unit test}
    \label{fig:ses-unit-test}
\end{figure}

Smart contracts are programs that will run on the blockchain, which means that once deployed, they will be immutable. Any error or problem in the contract can be difficult or even impossible to fix, so it is important to ensure that the contract is working correctly before deploying it to the network \cite{Zheng2017}. For this reason, we implemented unit tests to verify that the \gls{SES} token functions behave as expected and that the variables are updated according to the contract specifications. Figure \ref{fig:ses-unit-test} presents a test of tokens transfer between accounts. 

\begin{figure}[ht]
    \centering
    \includegraphics[width=.8\textwidth]{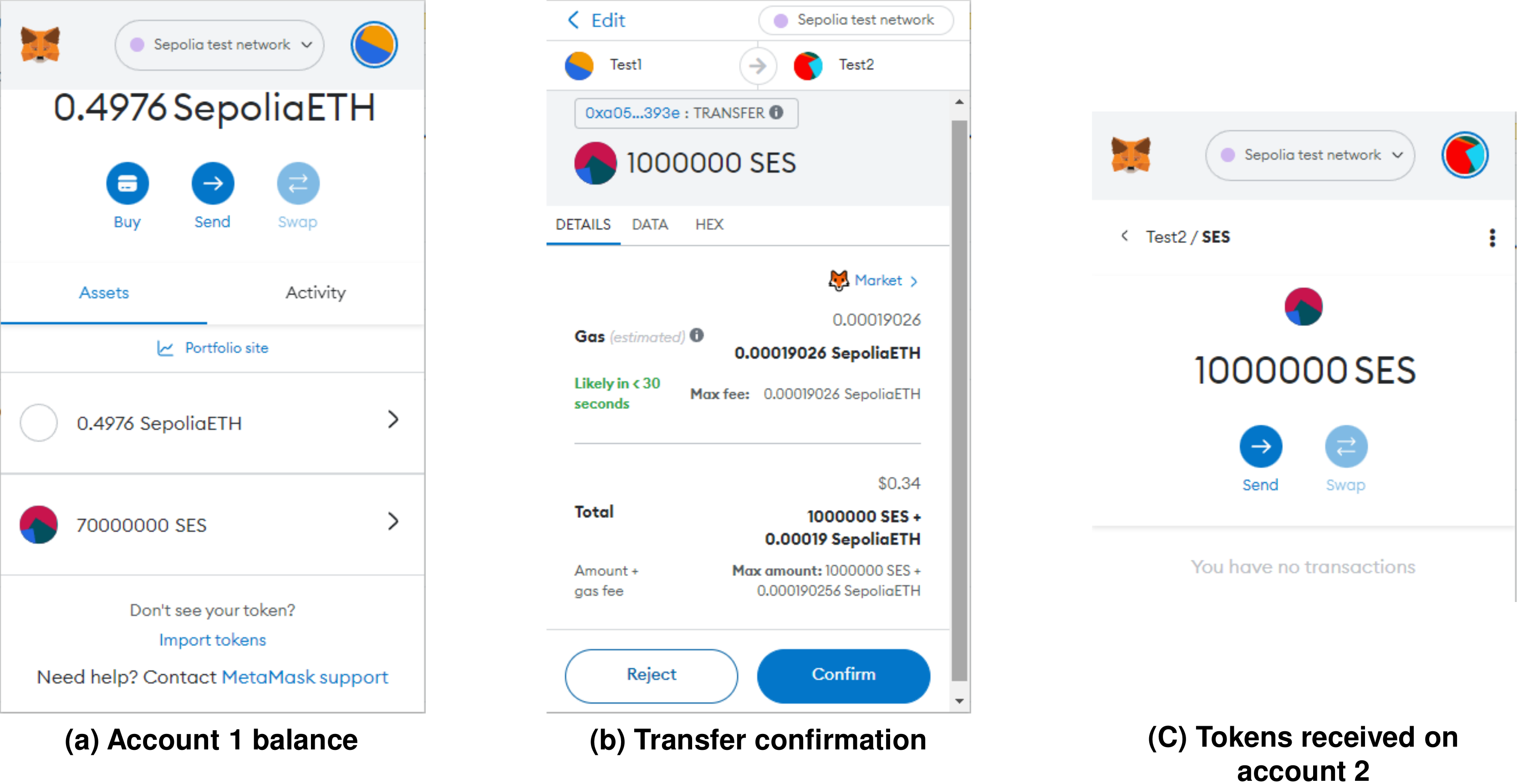}
    \caption{SES token transfer between test accounts on MetaMask wallet}
    \label{fig:metamask}
\end{figure}

After all the unit tests passed successfully, we performed a basic integration test with the MetaMask\footnote{https://metamask.io/} wallet. 
MetaMask is a browser extension that allows users to access \glspl{DAPP} based on blockchain, such as Ethereum, directly from their web browsers. It is a digital wallet that enables users to store, send, and receive cryptocurrencies, as well as interact with decentralized applications \cite{Zheng2017}.
Figure \ref{fig:metamask} presents the transfers between test accounts. 

Firstly, we added the Sepolia\footnote{https://sepolia.dev/} testnet in the wallet and integrated it with the \gls{SES} token. A testnet is a network in the blockchain ecosystem that allows developers and users to test and experiment with blockchain applications and smart contracts in a simulated environment that closely resembles the real blockchain network. From the \gls{SES} local token project on the desktop, 70,000,000 tokens were sent to test account 1, as shown in Figure \ref{fig:metamask} (a). According to Figure \ref{fig:metamask} (b), there is a transfer confirmation between accounts of 1,000,000 tokens. After the confirmation, these tokens are transferred to test account 2, as shown in Figure \ref{fig:metamask} (c).

\begin{figure}[ht]
    \centering
    \includegraphics[width=.7\textwidth]{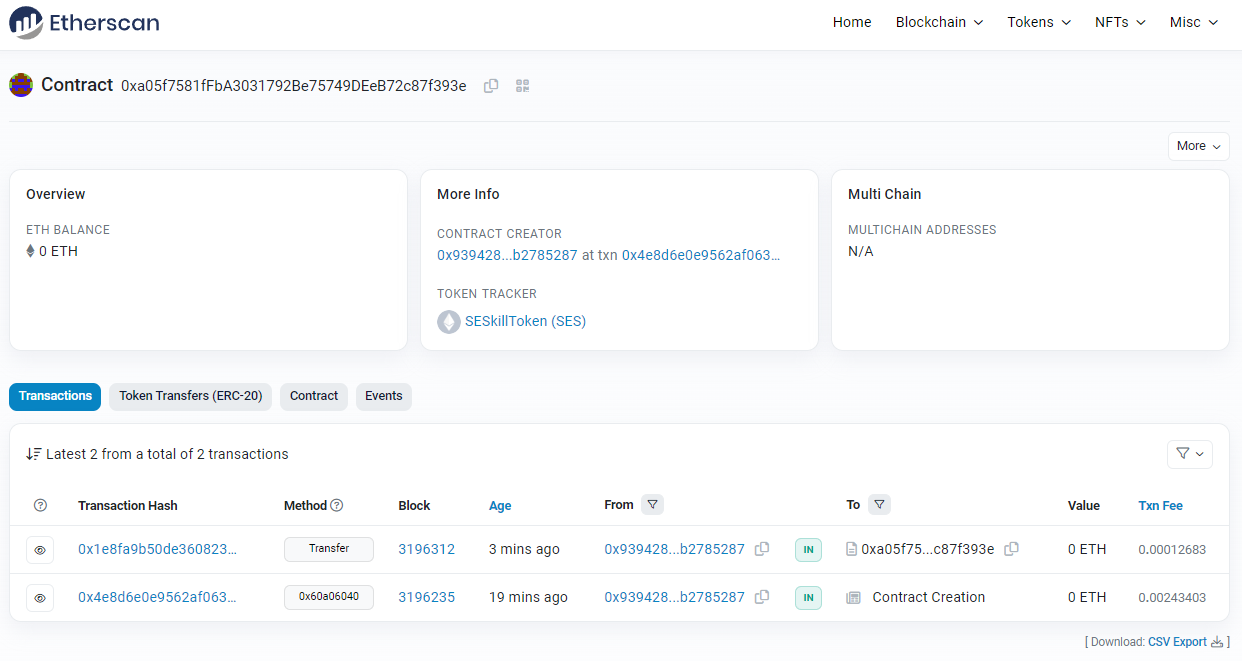}
    \caption{Transactions report on the Etherscan}
    \label{fig:etherscan}
\end{figure}

In order to prove these operations, these transactions were searched on Etherscan\footnote{https://etherscan.io/}. It is a popular blockchain explorer for the Ethereum network that allows users to view and search for transactions, addresses, and other activities on the blockchain. According to Figure \ref{fig:etherscan}, transfer transactions were performed successfully.

\section{Conclusion} \label{sex:conclusion}
This work described a token to support blockchain-based \gls{SEE}, named \gls{SES}. The implementation of \gls{SES} token-based rewards in education presents a range of advantages for learners and educators. \gls{SES} token provides a tangible incentive for learners to engage in educational activities and achieve learning outcomes and can create a more engaged and collaborative community of students, which can be an additional source of motivation for learning. 

Although the implementation of token-based rewards in education presents several advantages, it also entails several disadvantages that should be considered. One of the primary concerns is that students may become more focused on earning tokens rather than on the learning process itself, which can lead to a decrease in the quality and depth of learning. Another concern is the potential for token-based rewards to reinforce existing power structures and inequalities within the education system. For instance, learners who have greater access to resources or who are already performing well may be more likely to earn tokens, while those who are disadvantaged or struggling may be left behind.

As future works, we intend to perform research to solve token-based rewards concerns, the evolution of the \gls{SES} token implementation in order to integrate with the \gls{METASEE} platform, as well as evaluation with students, educators, and developers.

\bibliographystyle{sbc}
\bibliography{references}

\end{document}